# Hybrid, Optical and Wireless Near-Gigabit Communications System

L. Rakotondrainibe [1], Y. Kokar [1], G. Zaharia [1], G. El Zein [1], E. Tanguy [2], H. Li [2], B. Charbonnier [3]

[1] *IETR-INSA, UMR CNRS 6164, Rennes, France*
[2] *Université de Nantes, IREENA, France*
[3] *Orange Labs, Lannion, France*
[1] lrakoton@insa-rennes.fr

*Abstract*— **This paper presents the study and the realization of a hybrid 60 GHz wireless communications system. As the 60 GHz radio link operates only in a single-room configuration, an additional Radio over Fibre (RoF) link is used to ensure the communications in all the rooms of a residential environment. A single carrier architecture is adopted. The system uses low complexity baseband processing modules. A byte/frame synchronization technique is designed to provide a high value of the preamble detection probability and a very small value of the false alarm probability. Conventional RS (255, 239) encoder and decoder are used to correct errors in the transmission channel. Results of Bit Error Rate (BER) measurements are presented for various antennas configurations.**

## I. INTRODUCTION

The 60 GHz band, due to a large bandwidth is one of the most promising solutions to achieve a gigabit class for short distance high speed communications. Aspects including standardization, justification of using 60 GHz frequency, 60 GHz propagation, antennas and key system design issues were addressed in [1-4]. The characteristics of the 60 GHz band affect the system design. Problems such as power amplifier (PA) non-linearity and oscillator phase noise are more important for the circuits design. These effects should be taken into account in the overall communication system.

This paper proposes a hybrid optical/wireless system derived from simplified IEEE802.15.3c PHY layer proposal to ensure near 1 Gbps data rate on the air interface. The first system application in a point-to-point configuration is the high-speed file transfer. The system must operate in indoor, line-of-sight (LOS) domestic environments.

In sections II and III, the general design of the transmitter and the receiver are respectively described. In these sections, the intermediate frequency (IF) and radiofrequency (RF) blocks are first presented. Then, the baseband (BB) blocks are described. The byte/frame synchronization method is presented in baseband section. Measurement results are presented in section IV. Section V concludes the work.

## II. TRANSMITTER DESIGN

Fig. 1 shows the block diagram of the transmitter (Tx). The system uses a single carrier architecture based on a Differential encoded Binary Phase Shift Keying (DBPSK) modulation. A differential encoder is used before the BPSK modulator to remove the phase ambiguity at the receiver (Rx), knowing that a differential demodulation is used.

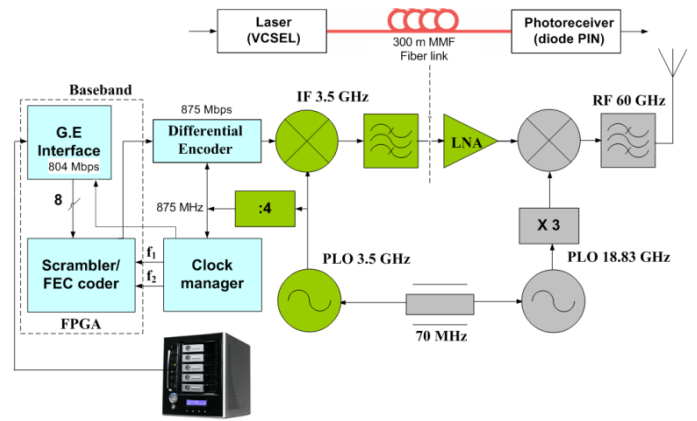

Fig. 1. 60 GHz wireless Gigabit Ethernet transmitter

### A. IF-Tx and RF-Tx architecture

After channel coding and scrambling, the input data are differentially encoded using logic circuits (PECL). The differential encoder performs the delayed modulo two addition of the input data with the output. The obtained data are used to modulate an IF carrier generated by a 3.5 GHz phase locked oscillator (PLO) with a 70 MHz external reference. The IF modulated signal is fed into a band-pass filter (BPF) with 2 GHz bandwidth, and transmitted through a 300 meters fibre link. This IF signal is used to modulate directly the current of the Vertical Cavity Surface Emitting Laser (VCSEL) operating at 850 nm through a band-pass filter with a bandwidth of 2 GHz. The VCSEL input RF power must not exceed -3 dBm to avoid signal distortions. After transmission, the optical signal is converted to an electrical signal by a PIN diode and amplified. The overall RoF link has 0 dB gain and an 8 GHz bandwidth. This bandwidth could be increased if necessary by using a VCSEL and a photodetector of broader bandwidth.

Following the RoF link, the IF signal is sent to the RF block. This block is composed of a mixer, a frequency tripler, a PLO at 18.83 GHz and a band-pass filter (59-61 GHz). The local oscillator frequency is obtained with an 18.83 GHz PLO with the same 70 MHz reference and a frequency tripler. The phase noise of the 18.83 GHz PLO signal is about –110 dBc/Hz at 10 kHz off-carrier. The BPF prevents spill-over into adjacent channels and removes out-of-band spurious signals caused by the modulator operation. The 0 dBm obtained signal is fed into the horn antenna with a gain of 22.4 dBi and a half-power beamwidth (HPBW) of 12°.

## B. BB-Tx architecture

The Gigabit-Ethernet (G.E.) interface is used to connect a home server to a wireless link with around 800 Mbps bit rate, as shown in Fig. 2. The Gigabit Media Independent Interface (GMII) is an interface between the Media Access Control (MAC) device and the physical layer (PHY). The interface defines speeds up to 1 Gbps, implemented using 8 bits data interface clocked synchronized at 125 MHz. However, this frequency is different from the clock sampling data (104.54 MHz) generated by a clock manager, as explained latter. These clocks asynchronism provide data packet loss. In order to avoid jitters transmission, a programmable circuit (FPGA) is used as part of buffers memory.

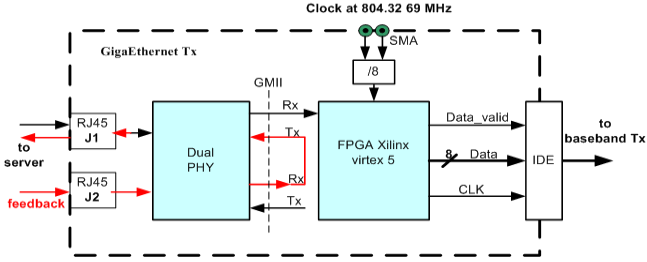

Fig. 2. Transmitter Gigabit Ethernet interface

The byte stream from the G.E. interface is transfered into the dual port FIFO memory (due to the RS encoder), as shown in Fig. 3. Otherwise, the data input must be paralellized by a serial-to-parallel (S/P) converter.

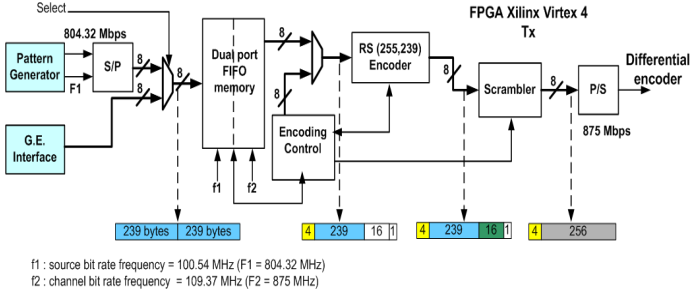

Fig. 3. BB-Tx architecture

The frame format consists of 4 preamble bytes, 239 data burst bytes, 16 check bytes and an additional dummy byte, as shown in Fig. 4. A known preamble is sent at the beginning of each frame to achieve the frame synchronization. The used preamble is a Pseudo-Noise (PN) sequence of 31 bits + 1 bit to provide 4 preamble bytes. A dummy byte should be added in the frame structure in order to make a code word of 256 bytes (a multiple of 4) useful for the scrambling operation.

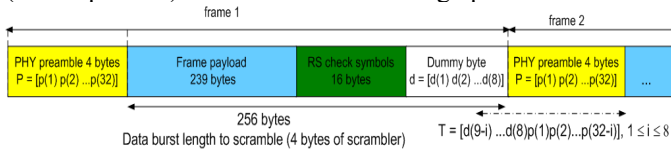

Fig. 4. Frame structure

Hence, owing to the frame structure, two different clock frequencies $f_1$ and $f_2$ are used in the baseband blocks:

$$f_1 = \frac{F_1}{8} = 100.54 \text{ MHz}, \quad f_2 = \frac{F_2}{8} = 109.37 \text{ MHz}.$$

where:

$$F_2 = \frac{IF}{4} = 875 \text{ MHz} \quad \text{and} \quad \frac{F_1}{F_2} = \frac{239}{260}.$$

The structure of the frame is obtained as follows: the input byte stream is written into the dual port FIFO memory at frequency $f_1$. The FIFO memory has been set up to use two different frequencies for writing at $f_1$ and reading at $f_2$. Therefore, reading can be started when the FIFO memory is half-full (to avoid overflow/underflow). The encoding control generates 4 preamble bytes and reads 239 data burst bytes (number of bytes used in the frame to be coded). The RS encoder reads one byte every clock period and bypasses the 4 preamble bytes. After 239 clock periods, the encoding control interrupts the bytes transfer during 17 triggered clock periods, so that 16 check symbol bytes and a dummy byte are added.

The dummy byte $d = [d(1)\ d(2)\ldots d(8)]$ is determined in order to obtain the lowest value of the maximum cross-correlation between the preamble P and T, where:

$$P = [P(1)\ P(2)\ \ldots..\ P(32)]$$

and

$$T = [d(9-i)\ \ldots\ d(8)\ P(1)\ P(2)\ \ldots P(32-i)],\ 1 \leq i \leq 8.$$

We note $k = 2^7*d(8)+2^6*d(7)+\ldots+ d(1)$, $0 \leq k \leq 255$. For each value of k, the maximum cross-correlation value between P and T is plotted in Fig. 5, where:

$$\text{Mcor}(k) = \max_i \left\{ \text{sum}(\overline{P \oplus T}) \right\}$$

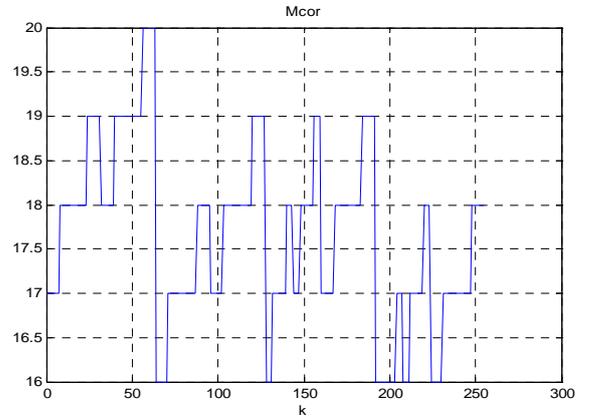

Fig. 5. Maximum cross-correlation between P and T for $0 \leq k \leq 255$

Consequently, the minimum of the maximum cross-correlation is obtained for several values of k. For each value of k, the cross-correlation between P and T is analyzed (for $1 \leq i \leq 8$) in order to identify the best dummy byte which gives the lowest cross-correlation values. The best result is obtained for k = 64 which gives d = [0 0 0 0 0 0 1 0]. As shown in Fig. 6, the maximum cross-correlation is equal to 16. This value is obtained twice; all the other values are smaller. This optimal dummy byte gives a smaller value of the preamble false detection probability.

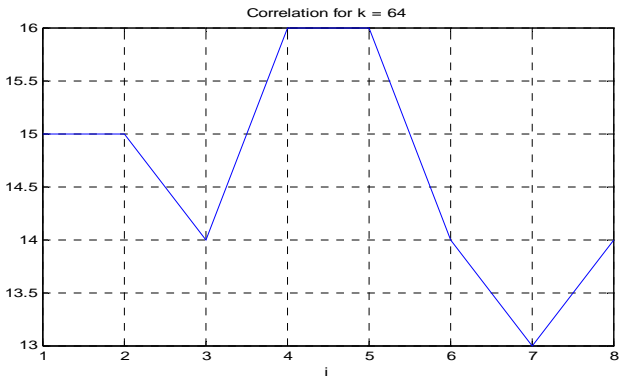

Fig. 6. Cross-correlation between P and T, for k = 64

The scrambler is a PN-sequence of 4 bytes (31 bits + 1 additional bit). This scrambler is chosen in order to provide the lowest cross-correlation values between the received scrambled data and the PHY preamble. This choice decreases the false alarm detection probability of the PHY preamble in the scrambled data. The obtained byte stream is finally serialized just before the differential encoder.

### III. RECEIVER DESIGN

Fig. 7 shows the receiver system block diagram. To simplify the receiver architecture, a non-coherent demodulation is used.

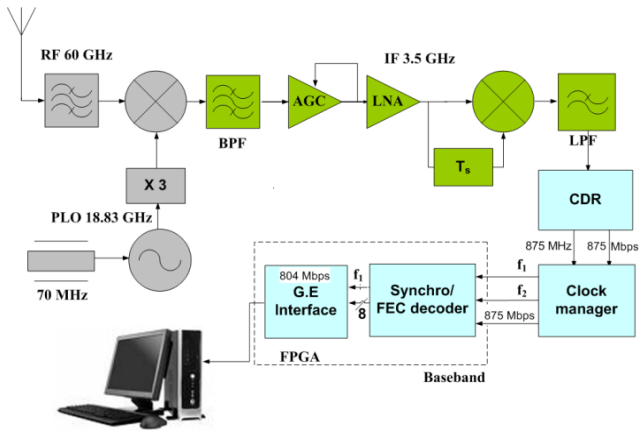

Fig. 7. 60 GHz Wireless Gigabit Ethernet receiver (Rx)

#### A. IF-Rx and RF-Rx architecture

The input band-pass filter removes the out-of-band noise. The RF signal at the output of the filter is down-converted to an IF signal centered at 3.5 GHz and fed into a band-pass filter with a bandwidth of 2 GHz. An Automatic Gain Control (AGC) with a dynamic range of 20 dB is used to ensure a quasi-constant signal level at the demodulator input. A Low Noise Amplifier (LNA) with a gain of 40 dB is used to achieve sufficient gain. A simple differential demodulation enables the coded signal to be demodulated and decoded. Compared to a coherent demodulation, this method is less performing in additive white Gaussian noise (AWGN) channel. The significant impact on the system caused by the radio channel is the frequency selectivity which induces inter-symbol interference (ISI). Nevertheless, the differential demodulation is more resistant to ISI effect. In fact, the differential demodulation, based on a mixer and a delay line (delay equal to the symbol duration $T_s$ = 1.14 ns), compares the signal phase of two consecutive symbols. Due to the product of two consecutive symbols, the rate between the main lobe and the second lobe of the impulse response of the channel is increased. Following the loop, a Low-Pass Filter (LPF) with 1 GHz cut-off frequency removes the high-frequency components of the obtained signal. For better acquisition of the Clock and Data Recovery (CDR) circuit, long sequences of '0' or '1' must be avoided. Therefore, the use of a scrambler and a descrambler are compulsory.

#### B. BB-Rx architecture

Fig. 8 shows the block diagram of the BB-Rx architecture. The synchronized data after the CDR are converted into a byte stream (due to the RS decoder).

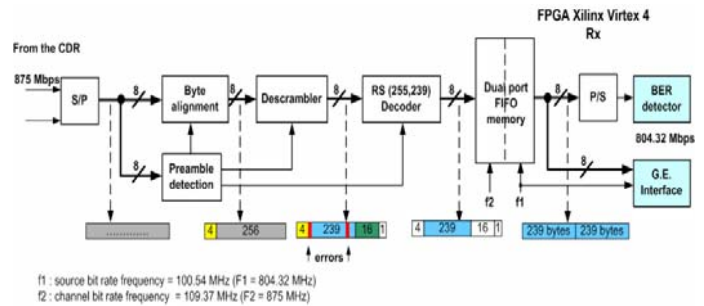

Fig. 8. BB-Rx architecture

Fig. 9 shows the architecture used for preamble detection and byte alignment.

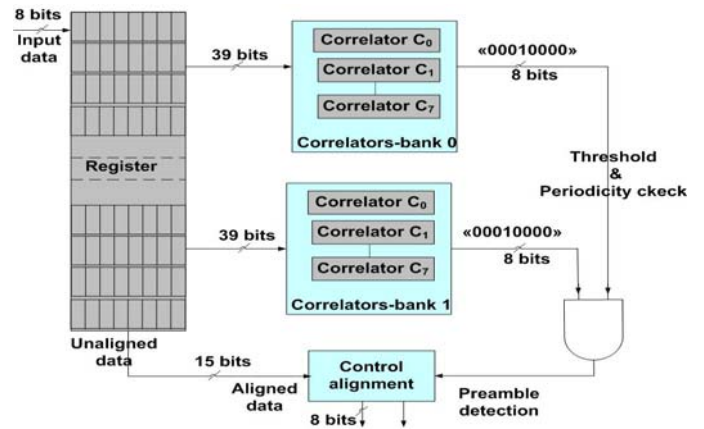

Fig. 9. Preamble detection based on periodicity and control alignment

The preamble detection is based on the cross-correlation between 32 successive received bits and the PHY preamble of 32 bits. Further, each correlator must analyze a 1-bit shifted sequence. Therefore, the preamble detection is performed with 32 + 7 = 39 bits (+ 7 because of different possible shifts of a byte). In all, there are 8 corrrelators of 32 bits in each correlators-bank, as shown in Fig. 9.

In order to minimize the preamble miss detection probability, two banks of correlators are used (taking into account the periodical repetition of the PHY preamble). The decision is made from 264 successive bytes (P1 + 256 data bytes + P2) stored in a register. Each value of the correlation between 32 successive received bits and the PHY preamble is compared to the threshold. The preamble is detected when the same correlator $C_k$ of each correlators-bank validated its presence.

The byte/frame synchronisation performance is characterized by the miss detection probability $P_m$, the false alarm probability $P_f$ and the channel error probability $p$ [5]. Fig. 10 shows the miss detection probability of the PHY preamble as a function of channel error probability $p$. The threshold $\gamma = 32$ is equal to the maximum cross-correlation value obtained when the received preamble has no error.

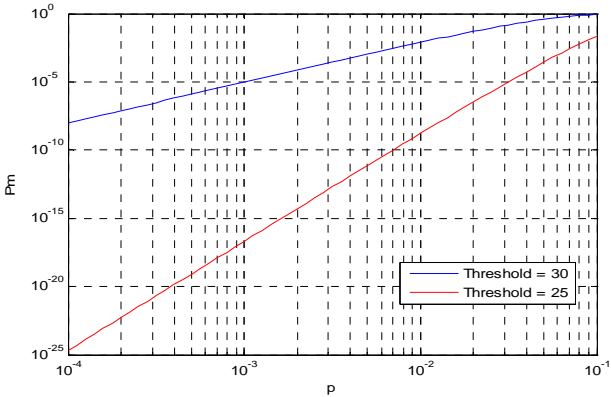

Fig. 10. Miss detection probability of PHY Preamble

The byte/frame synchronisation also depends on the false alarm events. Occasionally, due to the random values of the binary transmitted data, the architecture shown in Fig. 9 can falsely declare that the preamble is detected. A false alarm is declared when 2 correlators $C_k$ of the two correlator-banks indicate the detection of the preamble within the data burst bytes. Fig. 11 shows the false alarm probability $P_{F1}$ (with one correlators-bank) and $P_{F2}$ (with two correlators-banks) as functions of $p$.

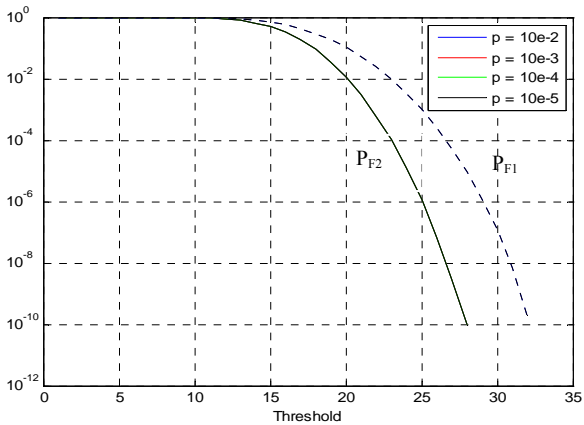

Fig. 11. False alarm probability of PHY Preamble

We can see that for $\gamma = 28$, $P_{F2} = 10^{-10}$ and $P_{F1} = 10^{-5}$. Therefore, by means of two correlators-banks, the best trade-off $\gamma = 28$ gives the high detection probability of the preamble and a very small false alarm probability. The effect of $p$ on the false alarm probability is insignificant since the random data bits "0" and "1" are considered equiprobable.

After the byte alignment and the preamble detection, the descrambler performs the modulo 2 addition between 4 successive received data and the scrambler sequence. The RS decoder processes the descrambled bytes and attempts to correct the eventual errors. The RS (255, 239) decoder can correct up to 8 erroneous bytes and operates at a high clock frequency (109.37 MHz). The byte stream obtained at the decoder output is written discontinuously into the dual port FIFO memory at the clock frequency $f_2$. A clock frequency $f_1 = 100.54$ MHz read out continuously the data bytes stored in the FIFO memory. Finally, the byte stream is transmitted to the receiver Gigabit Ethernet interface, as shown in Fig. 12.

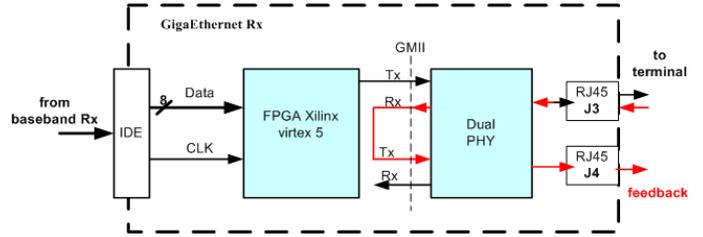

Fig. 12. Receiver Gigabit Ethernet interface

## IV. MEASUREMENT RESULTS

An HP 8753D vector network analyzer was used to determine the frequency response and the impulse response of RF blocks (Tx + Rx) including the LOS propagation channel. The objective was to measure the system bandwidth and to estimate the multipath channel effects. Measurements were located in a corridor where the major part of the transmitted power is focused to the receiver. The RF-Tx and RF-Rx were placed at a height of 1.5 m. After measurement set-up and calibration, 2 GHz available bandwidth was measured, as shown in Fig. 13.

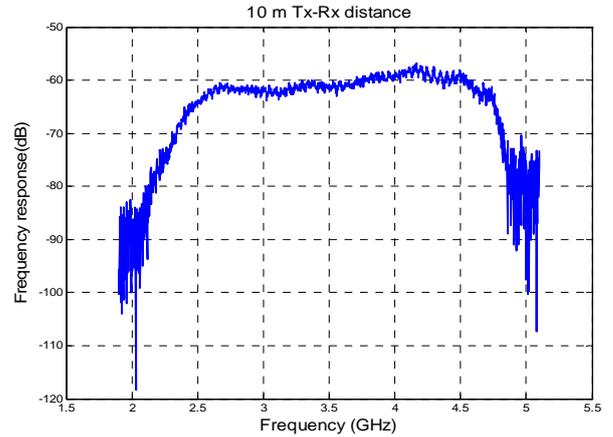

Fig. 13. Frequency response of RF Tx-Rx (with horn antennas)

A perfect system must have an impulse response with only one lobe. Fig. 14 presents the impulse response of the RF-Tx and RF-Rx blocks placed at 10 m distance. Few side lobes were obtained which are mainly due to RF components imperfections. A back-to-back test was realized by exchanging the Tx and Rx antennas with a 45 dB attenuator but similar results were obtained.

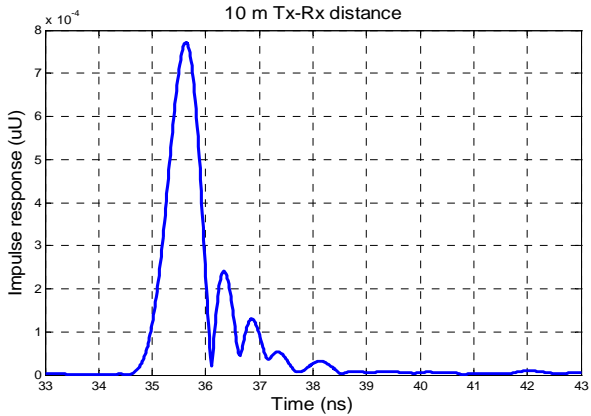

Fig. 14. Impulse response of RF blocks (using horn antennas)

Moreover, in order to evaluate the transmission quality, an HP70841B pattern generator was used to generate random data at the transmitter and an HP 708842B error detector was used at the receiver. Fig. 15 indicates a high transmission quality of the received data (good eye opening at 875 Mbps rate). As shown by this eye pattern, the recovered data can be obtained by sampling at half-period in order to be noise resistant. Fig. 16 shows the BER measurements at 875 Mbps as a function of the Tx-Rx distance. In our experiments, four antennas were used: two horn antennas and two patch antennas. Each patch antenna has an 8 dBi gain with a 30° at - 3 dB beamwidth.

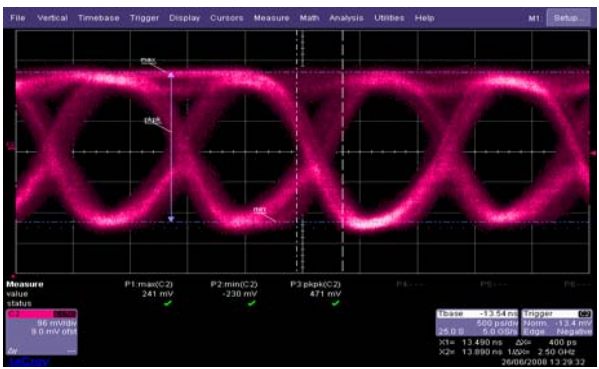

Fig. 15. Eye diagram at 875 Mbps (10 m Tx-Rx distance)

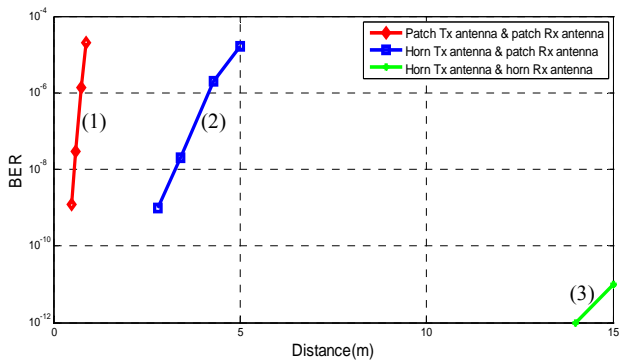

Fig. 16. BER performance at 875 Mbps (without channel coding)

The use of high-directivity horn antennas gives a remarkable BER performance and greatly reduces the harmful effects of multipath propagation. This leads to the saving of link budget as high as 28.8 dB compared to patch antennas. However, between (1) and (2), the link budget ratio is around of 15 dB. The system with high-gain directive antennas is acceptable for point-to-point (PP) communications links, with minimal multi-path interference. But, the 60 GHz radio links are sensitive to shadowing due to high attenuation of non-line-of sight (NLOS) propagation. In addition, the Tx and Rx antenna have to be aligned, otherwise the beam-pointing errors will cause an significant degradation of the channel quality.

For properly aligned antennas, if the direct path is blocked by moving objects, the communication can be completely lost.

In order to improve the system reliability, the choice of a centralized transmitter, preferably located on the centre of the room ceiling must be considered. The height of the transmitter (with less directional antenna) can partly reduce the influence of the shadowing. The receiver could be equipped with more directional antenna, pointed toward the transmitter.

If antennas beamwidth is large, equalization should be adopted to overcome multi-path interference while maintaining a high data rate. Futures works will provide BER (with channel coding) measurement in different environments.

## V. CONCLUSION

This paper presents the design and the implementation of a 60 GHz system for high data rate wireless communications. The proposed system provides a good trade-off between performance and complexity. An original method used for the byte synchronization is also described. This method allows a high preamble detection probability and a very small false detection probability. For 1 Gbps reliable communications within a large room, antennas must have a relatively high gain.

Increasing the data rate is achieved using higher order modulations such as QPSK. Equalization methods are still under study. The demonstrator will be further enhanced to prove the feasibility of wireless communications at data rates of several Gbps in different configurations, especially in non line-of-sight (NLOS) environments.

ACKNOWLEDGMENT

This work is part of research project Techim@ges supported by the French "Media & Network Cluster" and the Comidom project supported by the "Région Bretagne".